\documentclass[aps,prl,preprint,groupedaddress]{revtex4}
\usepackage{graphicx}
\usepackage{amsmath}

\begin{document}        

\title{ Search for Lepton Flavor Violation Process $e^+e^- \to e\mu$ 
        in the Energy Region $\sqrt{s}=984$ -- 1060 MeV and $\phi\to e\mu$
	decay}
\author{M.~N.~Achasov}
\email{achasov@inp.nsk.su}
\author{K.~I.~Beloborodov}
\author{A.~V.~Bergyugin}
\author{A.~G.~Bogdanchikov}
\author{\fbox{A.~D.~Bukin}}
\author{D.~A.~Bukin}
\author{T.~V.~Dimova}
\author{V.~P.~Druzhinin}
\author{V.~B.~Golubev}
\author{I.~A.~Koop}
\author{A.~A.~Korol}
\author{S.~V.~Koshuba}
\author{A.~P.~Lysenko}
\author{E.~V.~Pakhtusova}
\author{S.~I.~Serednyakov}
\author{Yu.~M.~Shatunov}
\author{Z.~K.~Silagadze}
\author{A.~N.~Skrinsky}
\author{A.~V.~Vasiljev} 
	 
\affiliation{Budker Institute of Nuclear Physics, 
	 Siberian Branch of the Russian Academy of Sciences 
	 11 Lavrentyev,Novosibirsk,630090, Russia}
\affiliation{Novosibirsk State University, 
	 630090, Novosibirsk, Russia}
\date{\today}

\begin{abstract}
The search for lepton-flavor-violation process $e^+e^-\to e\mu$ in the
energy region $\sqrt{s}=984$ -- 1060 MeV with SND detector at VEPP-2M
$e^+e^-$ collider is reported. The model independent 90\% CL upper limits 
on the $e^+e^-\to e\mu$ cross section, $\sigma_{e\mu} < 11$ pb, as well as on 
the corresponding $\phi\to e\mu$ branching fraction,  $B(\phi\to e\mu) < 2
\times 10^{-6}$, for the final particles polar angles $55^\circ<\theta<
125^\circ$, were obtained.
\end{abstract}


\maketitle

 For the most of fundamental fermions (quarks and neutrinos) the processes
 with flavor violation, quarks decays and neutrinos oscillation, are known.
 At the same time the LFV processes with charged leptons  has never been
 observed.  Theoretically  the processes of this kind are not strictly
 forbidden and can occur in many extensions of the Standard Model.

\begin{figure}[htp]
\includegraphics[height=5.0cm]{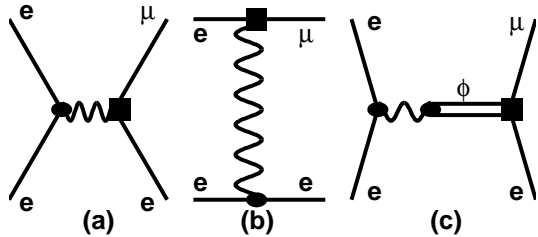}
\caption{\label{guarpa} The diagrams of the $e^+e^-\to e\mu$ process.}
\end{figure}

For the LFV hunting, the decays of $\mu$ and $\tau$ leptons, as well as
of the $Z$-boson and of various quark-antiquark mesons ($K,B,D,\eta,J/\psi,
\Upsilon$), along with a conversion process $\mu N \to eN$ are 
used \cite{Marciano,obzor}. The annihilation processes $e^+e^-\to e\mu$, 
$e\tau$, $\mu\tau$ are also suitable for this purpose. Theoretically 
these processes and related gauge boson and vector meson decays were studied, 
for example, in \cite{teor}. On the experimental side, the searches for the decays 
$J/\psi\to e\mu,e\tau,\mu\tau$ \cite{bes}, $\Upsilon\to\mu\tau$ \cite{cleo},
$Z\to e\mu,e\tau,\mu\tau$ \cite{lep}, as well as for the annihilation processes 
$e^+e^-\to e\tau$, $\mu\tau$ in the $\Upsilon(4S)$ energy domain 
\cite{babar},  and for the processes $e^+e^-\to e\mu$, $e\tau$, $\mu\tau$ in the 
energy region $\sqrt{s}=189$ -- $209$ GeV \cite{opal} were performed. 
However, in the energy region below the $J/\psi$ production threshold such 
studies were not done yet. In the $\phi(1020)$-meson energy domain, it is 
possible to search for the LFV process $e^+e^-\to e\mu$ and the corresponding
decay $\phi\to e\mu$ (Fig.\ref{guarpa}).

Existing stringent bounds on LFV $\mu\to 3e$ decay can be transformed to a 
severe constraint on the two-body $\phi\to e\mu$ branching fraction: 
$B(\phi\to e\mu)\le 4\times 10^{-17}$ unless some magic cancellations take
place in the  $\mu\to 3e$ decay amplitude \cite{Nussinov}. At first sight,
such a strong constraint makes doubtful any experimental effort to search
this decay. However, the magic cancellations mentioned above, although
unlikely, cannot be absolutely excluded. 

This work reports the results of studies of the process $e^+e^-\to e\mu$ 
in the energy region $\sqrt{s}\sim 1$ GeV with SND detector at $e^+e^-$
collider VEPP-2M.

The SND detector \cite{sndnim} operated from 1995 to 2000 at the
VEPP-2M \cite{vepp2} collider in the energy range $\sqrt[]{s}$ from 360 to
1400 MeV. The detector contains several subsystems. The tracking system
includes two cylindrical drift chambers. The three-layer spherical
electromagnetic calorimeter is based on NaI(Tl) crystals.
The muon system consists of plastic scintillation counters and two
layers of streamer tubes. The calorimeter energy and angular resolutions
depend on the photon energy as
$\sigma_E/E(\%) = {4.2\% / \sqrt[4]{E(\mbox{GeV})}}$ and
$\sigma_{\phi,\,\theta} = {0.82^\circ / \sqrt[]{E(\mathrm{GeV})}} \oplus
0.63^\circ$. The tracking system angular resolution is about
$0.5^\circ$ and $2^\circ$ for azimuthal and polar angles respectively.

This work is based on the data collected in the scans of the $\phi$-meson 
energy region. The total integrated luminosity used is $IL=8.5$ pb$^{-1}$. 
The luminosity was measured using the process $e^+e^-\to e^+e^-$ with the
accuracy of about 2\%.
 
In the reaction $e^+e^-\to e\mu$ the final particles are detected by the 
tracking system and have substantively  different energy depositions in 
the calorimeter. The muon system detects muons with a probability of greater 
than 90\%, while electrons are detected by this system with the probability
of less than  0.2\%. To search for $e^+e^-\to e\mu$ process, the so called 
collinear events 
containing two charged particles were used. We assume that the charged 
particle with higher energy deposition in the calorimeter (particle number 
one) is an electron, while the particle with lower energy deposition (particle
number two) is a muon. The events were selected using the following criteria 
(subscripts 1 and 2 denote the particle number):
\begin{enumerate}
\item
$N_{cha}=2$, where $N_{cha}$ is the number of the charged particles
originated from the interaction point: $|z_{1,2}|<10$ cm and $r_{1,2}<1$ cm. 
Here $z$ is the coordinate of the charged particle production point along the 
beam axis (the longitudinal size of the interaction region $\sigma_z$ about 
2.5 cm), $r$  is the distance between the charged particle 
track and the beam axis in the $r-\phi$ plane;
\item
$|\Delta\theta|=|180^\circ-(\theta_1+\theta_2)|<20^\circ$, where $\theta$ is
the particle polar angle;
\item
 $|\Delta\phi|=|180^\circ-|\phi_1-\phi_2||<5^\circ$, where $\phi$ is the
 particle azimuthal angle;
\item
 $55^\circ<\theta_{1,2}<125^\circ$;
\item
the  angular region $240^\circ<\phi_{1,2}<300^\circ$ not covered with the
muon system was excluded;
\item
the muon system was hited by the second particle and was not hited by the 
first one;
\item
$20<E_2^I<50$ MeV, $40<E_2^{II}<80$ MeV and $50<E_2^{III}<90$ MeV, where
$E_i^j$ are the energy depositions in the calorimeter layers, $i$ denotes the 
particle number and $j=I,II,III$ is the layer number;
\item
$E_1^I>70$ MeV, $E_1^{II}>130$ MeV and $20<E_1^{III}<100$ MeV.
\end{enumerate}
 
As a result 146 events were selected. The visible cross section (the events
number divided by the integrated luminosity) varies weakly with beam energy.
No contribution from the $\phi$-meson decays $\phi\to K^+K^-$, $K_SK_L$, 
$\pi^+\pi^-\pi^0$ is seen. This agrees with the expectations from the 
Monte-Carlo (MC) simulation. The events from the background process 
$e^+e^-\to\pi^+\pi^-$ can pass the selection if one of the pions looses its 
energy due to the ionization, while the other pion --  due to  nuclear
interactions. 
 
\begin{figure}[htp]
\includegraphics[width=13cm]{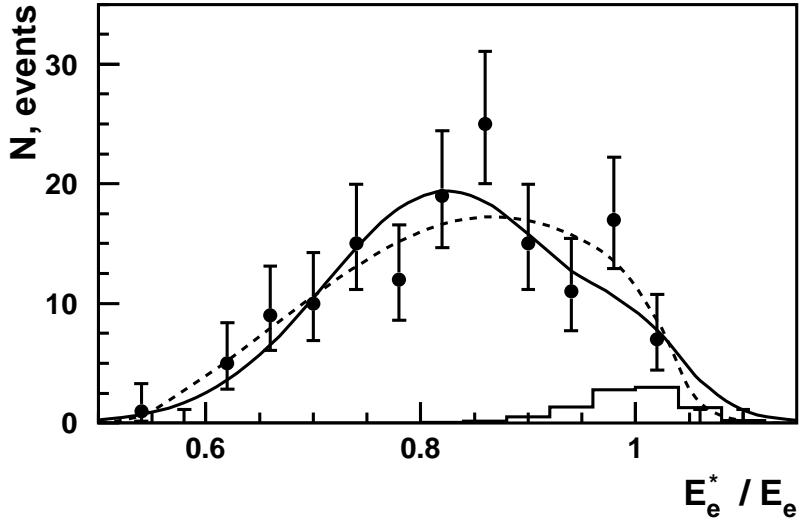}
\caption{The first particle $E^*_e/E_e$ for selected events (dots with errors).
         Solid curve -- fit by a sum of the distribution for electrons 
         (histogram) and Gaussian. Dashed curve  -- fit by a sum of the 
         distribution for electrons (histogram) and third-order polynomial.}
\label{el1n-nog}
\end{figure}
In order to obtain the cross section of the process $e^+e^-\to e\mu$ in the
whole energy region $\sqrt{s}=984$ -- 1060 MeV, the $E_e^*/E_e$ distribution
(Fig.\ref{el1n-nog}) was analyzed. Here $E_e^*$ and $E_e$ are the electron 
energies measured by the calorimeter and expected from the process kinematics 
respectively. To obtain the number of $e^+e^-\to e\mu$ events ($N_{e\mu}$), 
the $E_e^*/E_e$ spectrum was fit by a sum of distributions for electrons and
background. The distribution for electrons was obtained using experimental
data. The background was approximated either by Gaussian function or by a 
third-order polynomial. The coefficients of the background
function and $N_{e\mu}$ were free parameters of the fit. When the 
background was approximated with Gaussian, it was found that
$$
 N_{e\mu} = 12 \pm^{14}_{16}.
$$
This corresponds to the upper limit
$$
  N_{e\mu} < 30 \mbox{~~CL=90\%}.
$$
In the case of the third-order polynomial it was obtained
$$
 N_{e\mu} = 7 \pm^{11}_9.
$$
The corresponding upper limit is
$$
 N_{e\mu} < 21 \mbox{~~CL=90\%}.
$$
The higher limit $N_{e\mu} < 30$ was used for the further considerations.

Tracking system detection efficiency for the $e^+e^-\to e\mu$ events,
$\varepsilon_{track}$ (cuts 1 -- 5), was  obtained from MC simulation 
\cite{sndnim,sndpi2}. The MC events were generated with $1+cos^2\theta$ 
distribution.  The detection efficiency obtained for the angular
region $55^\circ<\theta<125^\circ$ actually  does not depend on the model of
the $\theta$-distribution. It's equal to $\varepsilon_{track}=0.59$. The 
experimental and simulated $\theta$,  $\Delta\theta$ and $\Delta\phi$ 
distributions for the processes $e^+e^-\to e^+e^-$, $e^+e^-\to\pi^+\pi^-$,  
$e^+e^-\to\mu^+\mu^-$ are in a good agreement \cite{sndpi2,sndmu2}. The 
systematic uncertainty associated with $\varepsilon_{track}$ determination is 
estimated to be less than 3 \%.

The efficiencies of muon and electron detection by the muon system were 
obtained using $e^+e^-\to\mu^+\mu^-$ and $e^+e^-\to e^+e^-$ data events. 
The $e^+e^-\to\mu^+\mu^-$ events  were selected according to the 
criteria 1--5 described above. The additional cut $r_{1,2}<0.1$ cm was used for
suppression of the cosmic ray background. The cut 7 was imposed on 
both particles. One particle was required to hit the muon system, while the 
other particle was used to determine the detection efficiency. The residual 
cosmic background was subtracted using the distribution of the
$z$-coordinate of the particles production point \cite{sndmu2}. Detection 
efficiency $\varepsilon^\mu_{muon}$ depends on the muon energy. Its value 
varies from 0.90 to 0.95 with the average value of $\varepsilon^\mu_{muon}$ 
being equal to 0.94. The $e^+e^-\to e^+e^-$ events were selected by the cuts 
1-- 5 and the condition $E_{1,2}/E_0> 0.7$. It was found that 
$1-\varepsilon^e_{muon}=0.998$.
  
Probabilities for muons and electrons to pass condition on the energy 
deposition in the calorimeter were obtained in a similar way. 
The $e^+e^-\to\mu^+\mu^-$ events were selected using cuts 1 -- 5 and
additional requirements $r_{1,2}<0.2$ cm, $E_{1,2}/E_0<0.6$. It was required 
that the muon system was hit by the both particles. The cosmic background was
suppressed by the restriction $|\tau_1-\tau_2|<5$ ns, where $\tau_{1,2}$ are 
time intervals between the signals from the scintillation counters and the beam
collision moment. The probability for muons to pass cut 7 was found to be
$\varepsilon^\mu_{cal}=0.86$.  The $e^+e^-\to e^+e^-$ events were selected
using criteria  1 -- 5 and requiring that the muon system was not fired by 
any particle and that the energy deposition of a randomly chosen particle was 
greater than $0.85\times E_0$. Other particle was used to obtain probability 
to satisfy the criterion 8, $\varepsilon^e_{cal}=0.70$. The values of 
$\varepsilon^\mu_{cal}$ and $\varepsilon^e_{cal}$ do not depend on the beam
energy.
 
\begin{figure}[htp]
\includegraphics[width=13cm]{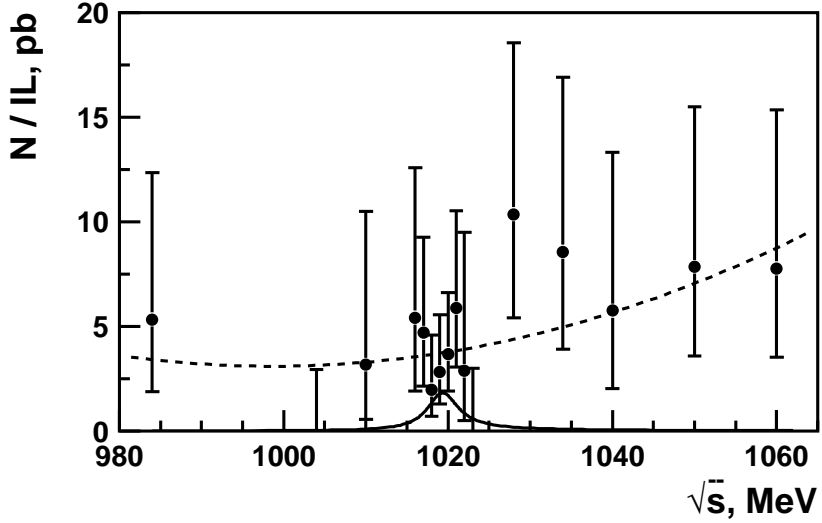}
\caption{The visible cross section obtained after additional cut 
         $0.9<E^*_e/E_e<1.1$. The solid curve is expected resonance line shape 
         corresponding to the upper limit on $B(\phi\to e\mu)$, 
         dashed curve is background approximation.}
\label{bug-ce-2}
\end{figure}
The detection efficiency of the  $e^+e^-\to e\mu$ process was calculated as
follows
$$
 \varepsilon_{e\mu}=\varepsilon_{track}\times
             \varepsilon^\mu_{muon}\times
             (1-\varepsilon^e_{muon})\times
             \varepsilon^\mu_{cal}\times
             \varepsilon^e_{cal}
$$
and the $e^+e^-\to e\mu$ cross section as
$$
 \sigma_{e\mu} = \frac{N_{e\mu}}{IL\varepsilon_{e\mu}},
$$
where $N_{e\mu}<30$, $\varepsilon_{e\mu}=0.31$ (the average  value 
$\varepsilon^\mu_{muon}=0.94$ was used), $IL=8.5$ pb$^{-1}$.  The following 
upper limit for the angular region $55^\circ<\theta<125^\circ$ was obtained
$$
\sigma_{e\mu} < 11 \mbox{~pb} \mbox{~~CL=90\%}.
$$
The upper limit on the $\phi\to e\mu$ decay was obtained assuming absence of
any non-resonance contribution and by using the additional cut 
$0.9<E^*_e/E_0<1.1$ (then $\varepsilon^e_{cal}=0.64$). The energy dependence
of the visible cross 
section is shown in Fig.\ref{bug-ce-2}. It was fit by the function:
\begin{eqnarray}
 \sigma = \varepsilon_{e\mu} \times (1+\delta_{rad}) \times 
 \frac{4\pi\alpha^2}{3s}
 \Biggl| \frac{3}{\alpha}
 {{\sqrt{B(\phi\to e^+e^-)B(\phi\to e\mu)}\frac{m_\phi\Gamma_\phi}
 {m_\phi^2-s-i\sqrt{s}\Gamma_\phi(s)}} }\Biggr|^2 + P_2(s),
\end{eqnarray}
where $(1+\delta_{rad})$ is the radiative correction factor \cite{KuraevFadin},
$P_2(s)$ is a second-order polynomial describing the background, 
$m_\phi$, $\Gamma_\phi$ are the $\phi$-meson mass and total width, 
respectively. The branching ratio and the coefficients of the $P_2(s)$ were 
free fit parameters. For the angular region $55^\circ<\theta<125^\circ$, it 
was obtained
$$
B(\phi\to e\mu) = (0.0\pm 1.5)\times 10^{-6},
$$
which corresponds to the upper limit
$$
B(\phi\to e\mu) < 2 \times 10^{-6}  \mbox{~~CL=90\%}.
$$
The presented upper limits do not depend on the angular distribution of the 
process $e^+e^-\to e\mu$.

The work is supported in part by RF Presidential Grant for Sc. Sch.
NSh-5655.2008.2, and by RFBR grants 08-02-00328-a, 08-02-00634-a, 
08-02-00660-a.

\end{document}